\documentclass{pasa}%

\usepackage{graphicx,url,amssymb,amsmath,rotating,color,units,wasysym,epsfig,multirow,epstopdf}
\usepackage[colorlinks,urlcolor=blue,citecolor=blue,linkcolor=blue]{hyperref}

\jid{PASA}
\doi{10.1017/pas.\the\year.xxx}
\jyear{\the\year}

\usepackage{aas_macros}

\usepackage{amssymb}
\usepackage{amsmath}
\usepackage{graphicx}
\usepackage{dcolumn}
\usepackage{color,units}
\usepackage{xspace}
\usepackage{mathtools}
\usepackage{tensor}
\usepackage{bm}
\usepackage{lipsum}
\usepackage{revsymb}
\usepackage[normalem]{ulem}

\newcommand{\Z}{\mathcal{Z}}
\newcommand{\M}{\mathcal{M}}
\newcommand{\V}{\mathcal{V}}
\renewcommand{\L}{\mathcal{L}}

\newcommand{\citeg}[1]{\citep[e.g.,][]{#1}}
\newcommand{\grbgw}[1]{\unit[2^{+10}_{-1}]{yr^{-1}}}
\newcommand{\jackpot}[1]{\unit[3^{+11}_{-2}]{yr^{-1}}}
\newcommand{\grbag}[1]{\unit[3^{+11}_{-2}]{yr^{-1}}}
\newcommand{\knonly}[1]{\unit[13^{+23}_{-10}]{yr^{-1}}}
\newcommand{\knag}[1]{\unit[4^{+18}_{-3}]{yr^{-1}}}
\newcommand{\coincident}[1]{\unit[14^{+25}_{-11}]{yr^{-1}}}
\newcommand{\coincidentpercent}[1]{41\pm 3\%}

\newcommand{\N}[1]{\mathcal{N}}

\newcommand{\SPA}{School of Physics and Astronomy, Monash University, Clayton VIC 3800, Australia}
\newcommand{\OzGravMonash}{OzGrav: The ARC Centre of Excellence for Gravitational Wave Discovery, Clayton VIC 3800, Australia}

\title[Multimessenger NEMO]{Multimessenger astronomy with a kHz-band gravitational-wave observatory\vspace{-1.3cm}}

\author[Sarin \& Lasky]{Nikhil Sarin$^{1,2,3,4}$\thanks{nsarin.astro@gmail.com} and Paul D. Lasky$^{1,2}$\thanks{paul.lasky@monash.edu}
\affil{$^1$\SPA}
\affil{$^2$\OzGravMonash}
\affil{$^3$Nordita, KTH Royal Institute of Technology and Stockholm University Roslagstullsbacken 23, SE-106 91 Stockholm, Sweden}
\affil{$^4$The Oskar Klein Centre, Department of Physics, Stockholm University, AlbaNova, SE-106 91 Stockholm, Sweden}
}

\begin{document}
\begin{frontmatter}
\maketitle



\begin{abstract}
Proposed next-generation networks of gravitational-wave observatories include dedicated kilohertz instruments that target neutron star science, such as the proposed Neutron Star Extreme Matter Observatory, NEMO. The original proposal for NEMO highlighted the need for it to exist in a network of gravitational-wave observatories to ensure detection confidence and sky localisation of sources. We show that NEMO-like observatories have significant utility on their own as coincident electromagnetic observations can provide the detection significance and sky localisation. We show that, with a single NEMO-like detector and expected electromagnetic observatories in the late 2020s and early 2030s such as the Vera C. Rubin observatory and SVOM, approximately $40\%$ of all binary neutron star mergers detected with gravitational waves could be confidently identified as coincident multimessenger detections. We show that we expect $\grbgw{}$ coincident observations of gravitational-wave mergers with gamma-ray burst prompt emission, $\knonly{}$ detections with kilonova observations, and $\knag{}$ with broadband afterglows and kilonovae, where the uncertainties are $90\%$ confidence intervals arising from uncertainty in current merger-rate estimates. Combined, this implies a coincident detection rate of $\coincident{}$ out to $\unit[300]{Mpc}$. These numbers indicate significant science potential for a single kilohertz gravitational-wave detector operating without a global network of other gravitational-wave observatories.
\end{abstract}
\end{frontmatter}

\section{Introduction}
Multimessenger gravitational-wave astronomy is a new field that already boasts a wealth of new scientific achievements, despite only having one detected event to date. The future of this field relies on continued improvement in both gravitational-wave and traditional astronomical instruments. Third-generation gravitational-wave observatories such as the Einstein Telescope~\citep{ET} and Cosmic Explorer~\citep{CE} are billion-dollar scale instruments with approximately a factor ten improvement in sensitivity over current, second-generation observatories like Advanced LIGO~\citep{LIGO}, Advanced Virgo~\citep{virgo}, and KAGRA~\citep{kagra}. Slated for the mid-2030s, these third-generation observatories still require significant technology development to see them realised. To this end, so-called 2.5-generation observatories have been proposed that serve as both dedicated astronomy observatories as well as technology drivers and facilitators for third-generation observatories. These 2.5-generation observatories include broadband interferometers such as A+~\citep{Aplus, Abbott2018} and Voyager~\citep{Voyager}, as well as dedicated kilohertz observatories~\citep{martynov19} such as the newly-proposed Neutron Star Extreme Matter Observatory, NEMO~\citep{NEMO}.

The NEMO design proposal held neutron star physics from binary neutron star mergers as the primary science driver. It proposed comparable high-frequency ($\gtrsim\unit[1]{kHz}$) sensitivity to Einstein Telescope and Cosmic Explorer, sacrificing low-frequency sensitivity ($\lesssim\unit[500]{Hz}$) to significantly reduce costs and the requirement for new technologies. The original proposal argued that NEMO must co-exist with 2.5-generation instruments to maximise scientific impact---those instruments would provide source localisation with their superior low-frequency sensitivity. However, during the proposed period of operation (late 2020s/early 2030s), there may be significant periods of time when NEMO must operate without a global network of gravitational-wave observatories. In this work, we show that NEMO-like observatories can operate in isolation without a heterogeneous global array because of the presence of electromagnetic telescopes with survey and/or all-sky capabilities that will make routine detections of the electromagnetic counterparts of binary neutron star mergers out to relevant distances. 

The Vera C. Rubin observatory~\citep{vera_rubin} is an 8.4-m optical/UV telescope capable of detecting kilonovae associated with binary neutron star mergers. Providing they are in the correct hemisphere, Vera Rubin will detect 100\% of kilonovae using only $\unit[30]{s}$ exposures out to $\sim\unit[450]{Mpc}$~\citep{chen21}, approximately equivalent to the maximum horizon distance of NEMO~\citep{NEMO}. In optical/UV, Vera Rubin is the tip of the iceberg; other likely facilities with even better sensitivity include the James Webb Space Telescope~\citep{jwst} and Keck Wide-Field imager~\citep{kwfi} as well as the thirty-metre class of ground-based facilities such as the Giant Magellan Telescope~\citep[GMT;][]{gmt}, and the Thirty-Metre Telescope~\cite[TMT;][]{tmt}. These will be complemented with all-sky survey facilities including Zwicky Transient Facility~\citep[ZTF;][]{ztf_paper} and Pan-STARRS~\citep[][]{panstarrs}, capable of providing nightly all-sky surveys to observe transient kilonovae and gamma-ray burst afterglows. Observatories like Evryscope may also provide continuous all-sky coverage~\citep{evryscope}.

By the late 2020s there will likely be multiple successor x-ray and gamma-ray telescopes to the Neil Gehrel's \textit{Swift} Observatory~\citep{swift} and Chandra~\citep{chandra}, including proposed instruments such as ATHENA~\citep{athena} and Theseus~\citep{Theseus}, and currently operational or upcoming facilities like SVOM~\citep{svom} and eROSITA~\citep{erosita}. Current instruments have the capability to detect on-axis bursts well beyond the horizon of NEMO~\citeg{howell19}, and proposed missions will be more sensitive with equivalent or better sky coverage, and faster slew times. These instruments even offer the possibility of detecting binary neutron star pre-cursor emission~\citeg{Most2020, Sridhar2021, Ascenzi2021}, as well as potential emission from the central engine following the merger~\citeg{sarin21_review}. Afterglows in radio will be detectable by instruments like the Square Kilometre Array beyond the horizon distance of NEMO~\citep{Dobie2021}, which will facilitate very-long baseline interferometry follow-up to determine jet structure and orientation of some short gamma-ray bursts from binary neutron star mergers~\citep{Duque2019}. Observatories like the Cherenkov Telescope Array will also provide all-sky high energy gamma-ray survey capabilities~\citep{cta}.

With all of these detections across multiple wavebands and with high-frequency gravitational waves, the problem becomes: \textit{how confidently can we associate any two measurements as coming from the same source?} Without this association, we cannot fulfil the rich promise that multimessenger gravitational-wave science brings.  In this Paper, we calculate predictions for the rate of binary neutron star mergers NEMO alone can detect and that can be confidently associated with their various electromagnetic counterparts. We show the gamma-ray prompt emission will be confidently associated with the gravitational-wave emission at an overall rate of $\grbgw{}$, where the uncertainties are $90\%$ confidence intervals arising from uncertainty in current merger-rate estimates. Gravitational waves and kilonovae will be associated at $\knonly{}$ with broadband afterglows at $\knag{}$. Combined, this implies a coincident detection rate of $\coincident{}$ out to $\unit[300]{Mpc}$.

In this work, we discuss only neutron star binary mergers. The aLIGO/aVirgo network have already detected $2^{+2}_{-2}$ (100\% confidence interval) neutron star-black hole binaries~\citep{abbott20_gw190814,abbott21_gwtc2,abbott21_gwtc2.1,abbott21_nsbh}, however we note that each of these purported mergers have black hole masses and spins such that the neutron stars are not expected to be tidally disrupted, and therefore no electromagnetic counterparts are expected~\citeg{foucart20}. The rates of neutron star-black hole binary mergers with tidal disruption are therefore unknown~\citeg{sarin21_nsbh}, and we do not speculate further on the expected science outcomes of those systems in this work. We further note that NEMO-like observatories are potentially good for detecting gravitational-wave signals from supernovae~\citeg{Powell2020, Mezzacappa2020, Pan2021}, however the rates of gravitational-wave only, and coincident gravitational-wave and electromagnetic detections, are highly uncertain. Finally, NEMO-like observatories may also be able to detect nearly-monochromatic signals from millisecond pulsars, however there the gravitational-wave amplitude is highly uncertain~\citep[e.g., see][and references therein; although see~\citet{woan18} for evidence of a minimum ellipticity]{lasky15_review,haskell21}.

This paper is set out as follows. In Sec.~\ref{sec:2030s} we detail the salient aspects of future gravitational-wave and electromagnetic facilities for this study and describe the modelling of the gravitational-wave and electromagnetic signal.  In Sec.~\ref{sec:assoc} we provide the main results of the paper, using a Bayesian method for determining coincidences between multiple different observations which is presented in Appendix~\ref{sec:method}. Section~\ref{sec:assoc} focuses first on a hypothetical `jackpot' event (Sec.~\ref{sec:170817}) that is both on axis and nearby at~\unit[40]{Mpc}---the distance of the first gravitational-wave multimessenger event GW170817. In Sec.~\ref{sec:distance} and \ref{sec:inclination} we provide more realistic scenarios by looking at populations of events with varying distances and inclination angles to the observer's line of sight, respectively. In Sec.~\ref{sec:conclusion} we conclude by discussing the promising multimessenger science that can be learned from these future coincident detections.

\section{Observations in the 2030s}\label{sec:2030s}
\subsection{Gravitational-wave observations}\label{sec:GW}
In this work, we assume only a single ground-based gravitational-wave interferometer is operating around the globe with NEMO-like sensitivity~\citep{NEMO}. While we hope this situation never eventuates, we are mindful that there may be a data gap between the end-of-life stage of 2.5-generation interferometers such as A+ and Virgo+~\citep{Abbott2018}, and the beginning of third-generation observatories such as Einstein Telescope and Cosmic Explorer~\citep{ET, CE}. 

We simulate a realistic population of binary neutron star gravitational-wave signals at fixed distances but with the inclination angle $\iota$ drawn from a uniform in $\cos\iota$ distribution. We inject all signals with masses drawn from a Gaussian mixture model motivated by analysis of the mass distribution of all neutron stars in our Galaxy~\citep{alsing18}. This mass distribution mixture model is 
\begin{align}
    p(M)=\left(1-\epsilon\right)\N{}\left(\mu_1,\,\sigma_1\right)+\epsilon\N{}\left(\mu_2,\,\sigma_2\right),\label{eq:pm}
\end{align}
where $\N(\mu,\,\sigma)$ denotes a normal distribution of mean $\mu$ and standard deviation $\sigma$, $\mu_1=1.32M_\odot$ and $\sigma_1 = 0.11$, $\mu_2=1.80M_\odot$, $\sigma_2=0.21M_\odot$, and mixing fraction $\epsilon=0.35$. We inject these signals isotropically over the sky.

We inject these signals into a NEMO detector located in Gingin, near Perth, Australia, to determine the fraction of events detectable at a given distance. We define a gravitational-wave detection as having an optimal matched-filter signal-to-noise ratio greater than eight. This threshold is somewhat arbitrarily defined---it is approximately the single-detector threshold currently used for the LIGO/Virgo Collaborations---however a more nuanced signal-to-noise threshold would require a detailed understanding of the instrument's noise properties. Regardless, the precise numerical values for the fraction of detectable events is not important, but the overall approximate numbers are useful to understand for the context of this paper. We note that the number of events is $\propto 1/\textrm{SNR}^3$. 

In the top right panel of Fig.~\ref{fig:detectable_events}, we plot the the total number of detectable events in a single NEMO as a function of distance as the solid coloured band. We use the current best estimate for the binary neutron star merger event rate of $\unit[320^{+490}_{-240}]{Gpc^{-3} yr^{-1}}$~\citep{Abbott2021_03acatalog}, where the uncertainties are the 90\% confidence intervals, which are shown as the dashed curves in each panel of Fig.~\ref{fig:detectable_events}. This figure indicates that NEMO will detect 95\%, 65\% and 5\% of binary neutron star mergers at $\unit[40]{Mpc}$, $\unit[100]{Mpc}$, and $\unit[300]{Mpc}$, respectively.

\subsection{Electromagnetic observations}\label{sec:EM}
The 2030s will see the full promise of several current and proposed electromagnetic observatories with all-sky or survey capabilities. These observatories will enable routine and confident detections of kilonovae, short gamma-ray bursts, and their afterglows beyond the horizon distance of NEMO~\citeg{Theseus, vera_rubin}. These observatories may also enable detection of precursor electromagnetic emission from binary neutron star mergers as well as their post-merger central engines out to distances relevant to NEMO~\citep[see][and references therein]{Ascenzi2021,sarin21_review}. 

In this work, we focus on whether the electromagnetic counterpart of binary neutron star mergers are detectable, and how confidently such signals can be associated with their gravitational-wave counterpart. We consider an electromagnetic signal to be detectable if it is above a pre-defined threshold for each instrument (described below). For gamma-ray burst prompt emission, their afterglows, and kilonovae, these thresholds are relatively well estimated for various electromagnetic telescopes~\citeg{Coward2011, Kanner2012,Siellez2014}. However, the thresholds, rates, and rate of false positives for precursor emission and post-merger emission from the central engine are unknown~\citeg{Ascenzi2021, Sridhar2021}. We therefore only focus on the relatively well observed prompt gamma-ray emission, broadband afterglow, and kilonova.

The prompt flash of gamma rays in short gamma-ray bursts is typically followed by a long-lasting, broadband afterglow. While the mechanism responsible for producing the prompt gamma-ray emission is ill understood~\citeg{kumar15}, the broadband afterglow is relatively simple to understand. Most afterglows are believed to be from the interaction of a relativistic jet with the surrounding interstellar medium~\citep{sari98}. There are additional types of afterglows that we do not consider in detail here, such as those suggesting an additional contribution from a neutron star~\citeg{zhang01,rowlinson13, lu15, sarin19}.

The multi-wavelength observations of GRB170817A confirmed the long-held suspicion that gamma-ray burst jets are structured, i.e., the energy and Lorentz factor have a non-trivial angular dependence~\citep{troja17_xrays,alexander18,mooley18a,mooley18_superluminal}. While the true jet structure is unknown, several phenomenological models have been developed and fit to gamma-ray burst observations. One such commonly used model is the Gaussian structured jet~\citeg{lamb18, afterglowpy}
\begin{equation}\label{eq:jetstructre}
    E(\theta)= E_{0} e^{-\frac{\theta^{2}}{2 \theta_{c}^{2}}}. 
\end{equation}
Here, $E_{0}$ is the isotropic equivalent kinetic energy in the afterglow, $\theta$ is the angle from the jet axis, and $\theta_{c}$ is the half-opening angle of the ultra-relativistic core or, equivalently, the opening angle of the jet. None of these parameters are well constrained empirically, but multi-wavelength observations of GRB170817A and other gamma-ray bursts offer some clues. 

The ultra-relativistic jet emanating from a neutron star merger interacts with the surrounding interstellar medium accelerating a fraction of electrons, $\xi_{n}$, with some fraction of the total energy of the jet, $\epsilon_{e}$, and some fraction of the energy in the magnetic field, $\epsilon_b$. The radiation produced by these electrons is responsible for the observed broadband afterglow. This interaction is relatively easy to model, but computationally expensive. Fortunately, semi-analytic approximations make modeling these broadband afterglows tractable. We use \verb!afterglowpy!~\citep{afterglowpy} with the above jet structure profile to model our afterglow lightcurves. 

The prompt emission is even less understood than the broadband afterglow.
There is no robust generative model that allows predictions or modeling of the prompt emission energetics.
Observations of the off-axis prompt emission from GRB170817A~\citeg{matsumoto19, ioka19} further highlighted our lack of understanding of the angular structure~\citeg{howell19, ioka19}.
By the late 2020s/early 2030s, we will undoubtedly have more insight into prompt energetics and angular structure. 
We therefore plough forward in our endeavour, acknowledging these known unknowns. 

In light of GRB170817A, one common approach is to compute the peak flux at a given observer viewing angle using a similar structured jet profile used to describe the afterglow~\citeg{howell19, Salafia2019, Theseus}. Such analyses makes two assumptions: 1) the jet structure in the afterglow and prompt emission phase are the same, and 2) the jet is transparent to the gamma rays at all viewing angles.
Both these assumptions could be incorrect. Alternatively, one may use a range of estimated gamma-ray efficiencies for prompt emission inside the opening angle of the jet~\citeg{fong15}, and model the emission outside the jet as a cocoon-shock breakout~\citeg{gottleib18}. This gives a qualitatively similar answer for prompt-emission energetics outside the jet opening angle as the approach of~\citet{howell19,Salafia2019}. We therefore adopt the latter approach. By the 2030s, we may better understand the energetics of off-axis emission if we observe more multimessenger events like GW170817~\citep{Biscoveanu2020}. 

All of the above ignores complications surrounding jet-launching itself. Numerical simulations suggest that no neutron star remnant can launch a jet that can produce a short gamma-ray burst~\citep{murguia17}. However, this is in tension with x-ray afterglow observations of many short gamma-ray bursts~\citeg{rowlinson13}. In this work, we stick with numerical results, placing a constraint that a gamma-ray burst and subsequently the afterglow emission is only possible when the remnant mass is above the threshold to form a black hole i.e., $M_{\rm {rem}} \gtrsim 1.2 M_{\rm {TOV}}$, where $M_{\rm {TOV}}$ is the maximum non-rotating neutron star mass; a property of the nuclear equation of state. For each merger, we estimate the remnant mass $M_{\rm {rem}}$ using the relationship described in~\citet{gao16}. We then calculate whether that merger would produce a gamma-ray burst and afterglow marginalising over piecewise polytropic equations of state with $M_{\rm {TOV}}$ between $2.2-2.4 M_{\odot}$ motivated by current estimates~\citeg{ai20}. 

Beyond the prompt gamma-ray emission and broadband afterglow, r-process nucleosynthesis from the neutron-rich ejecta of a binary neutron star merger is expected to power a week-long thermal transient, i.e., a kilonova. This was likely first seen in GRB130603B~\citep{Tanvir2013, Berger2013}, but has since been studied in considerably more detail through multi-wavelength observations of AT2017gfo, the kilonova counterpart to GW170817~\citeg{evans17, smartt17, villar17}. 
There are more kilonova models in the literature than there are observations. Many models do a reasonable job of explaining the observations of AT2017gfo~\citep[see][for a quantitative model comparison]{Breschi2021}. Some models are predominantly phenomenological while others attempt to connect the progenitor's parameters to the resulting lightcurve. We use the latter class of models. In particular, we use the kilonova model from~\citet{metzger17} relating it to the intrinsic binary parameters through fits of the ejecta properties derived from numerical-relativity simulations~\citep{Dietrich2017} with a piecewise polytropic equation of state~\citep[using the software presented in][]{HernandezVivanco2020} and a baryonic mass relation~\citeg{Coughlin2017,Gao2020}.

The afterglow emission is broadband, and will be detectable with various electromagnetic telescopes and surveys at different wavelengths. However, for the purposes of detectability here, we focus specifically on the optical afterglow with the Vera Rubin Observatory. The $5\sigma$ magnitude threshold for single images in Vera Rubin in $r$-band is $24.5$ with an equivalent flux density of $\unit[4.79\times10^{-4}]{mJy}$, which requires a $\unit[30]{s}$ integration time~\citep{vera_rubin}. Therefore, we consider any optical afterglow signal with an $r$-band peak flux greater than the single-image threshold to be detectable. We use the same threshold for a kilonova signal. For gamma-ray prompt emission, we consider a gamma-ray burst to be detectable if the peak gamma-ray flux in the $\unit[30-150]{keV}$ regime is above $\unit[3 \times 10^{-8}]{erg~cm^{-2}~s^{-1}}$. This is comparable to the expected sensitivity of Theseus~\citep{Theseus}. We could also use Theseus to study the detectability of the x-ray afterglow, however the x-ray instrument proposed for Theseus does not have the wide-field survey capability of Vera Rubin and will not serendipitously detect as many afterglows.

\begin{figure*}
    \centering
    \includegraphics[width=0.75\textwidth]{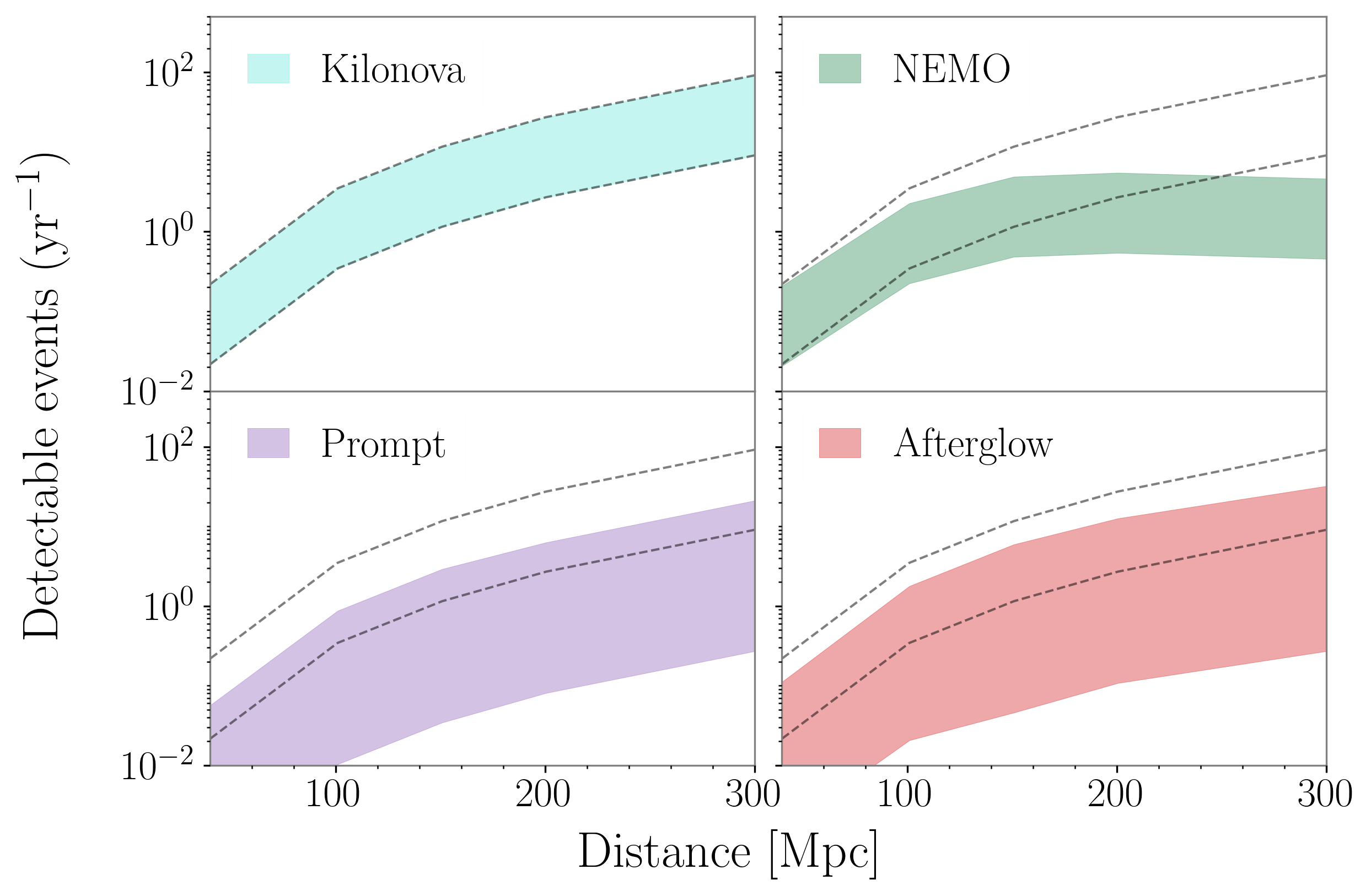}
    \caption{The rate of detectable events as a function of distance with different messengers. The dashed curves indicate the $90\%$ credible interval of the binary neutron star merger rate~\citep{Abbott2021_03acatalog}, while the coloured band represents the $90\%$ credible interval from marginalising over the model and systematic uncertainties for each messenger.}
    \label{fig:detectable_events} 
\end{figure*}

We simulate the expected electromagnetic emission using the models described above using the population of binary neutron star mergers described in Sec.~\ref{sec:GW}. The prior ranges of the various parameters in the relevant models are motivated by analyses of GRB170817A~\citep{afterglowpy} and AT2017gfo~\citep{Coughlin2017} and summarised in Table~\ref{tab:priors} in Appendix~\ref{appendix:priors}. We calculate the number of these events that produce a detectable electromagnetic counterpart at a given distance, which is shown in the top left panel of Fig.~\ref{fig:detectable_events} for kilonovae, bottom left panel for prompt emission, and bottom right panel for the broadband afterglows. The bands correspond to the uncertainties in detectable fraction due to the uncertain model parameters, nuclear equation of state, and binary neutron star merger rate. 
The absolute merger rate is shown in each panel as the dashed curves.

\section{Associating gravitational-wave and electromagnetic signals}\label{sec:assoc}
In the following we calculate coincident detection rates under different scenarios. We generalise the Bayesian calculation of coincidences between any two data sets first presented in~\citet{Ashton2018} to multiple data sets. We provide details of the formalism in Appendix~\ref{sec:method}.

\subsection{A GW170817-like event}\label{sec:170817}
We first walk through a fiducial example of a coincident multimessenger event akin to GW170817. We include a single NEMO-like gravitational-wave detector, with electromagnetic observations that provide prompt gamma-ray burst identification, broadband afterglow and kilonovae observations, and host-galaxy identification. This is truly a jackpot event; in subsequent sections we consider scenarios with only a subset of these observations. 

\begin{figure}
    \centering
    \includegraphics[width=0.45\textwidth]{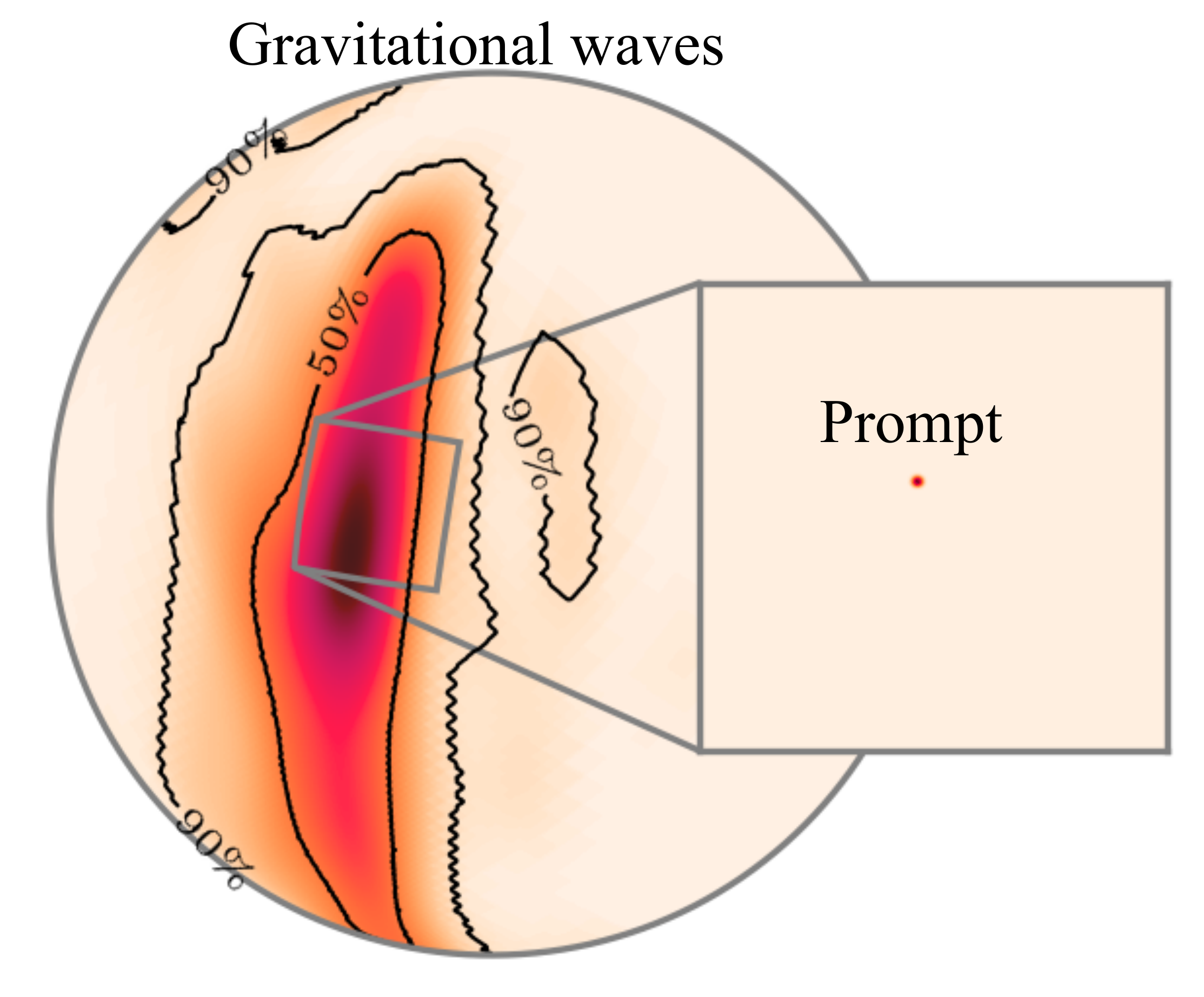}
    \caption{Sky localisation posterior distribution of GW301116, a GW170817-like on-axis binary neutron star merger at \unit[40]{Mpc}. The main plot shows the sky localisation posterior from a single gravitational-wave detector, NEMO. The inset shows the sky localisation posterior for the same event but from the detection of the prompt gamma-ray emission form a Theseus-like observatory.}
    \label{fig:sky} 
\end{figure}

\begin{figure}
    \centering
    \includegraphics[width=0.45\textwidth]{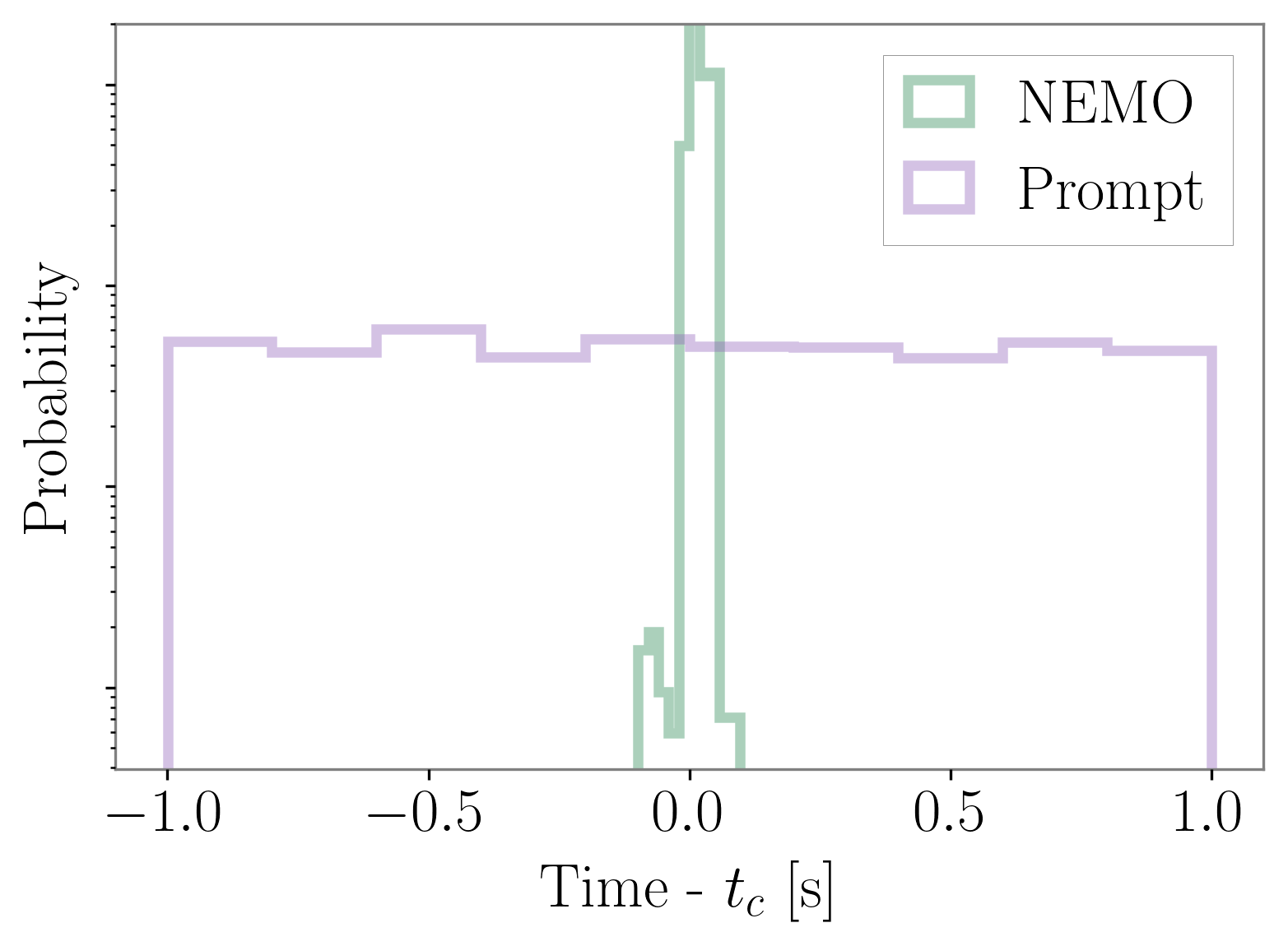}
    \caption{Posterior distribution for the merger time, $t_c$ for GW301116 from NEMO (green) and the prompt emission (purple).}
    \label{fig:tc} 
\end{figure}

\begin{figure}
    \centering
    \includegraphics[width=0.45\textwidth]{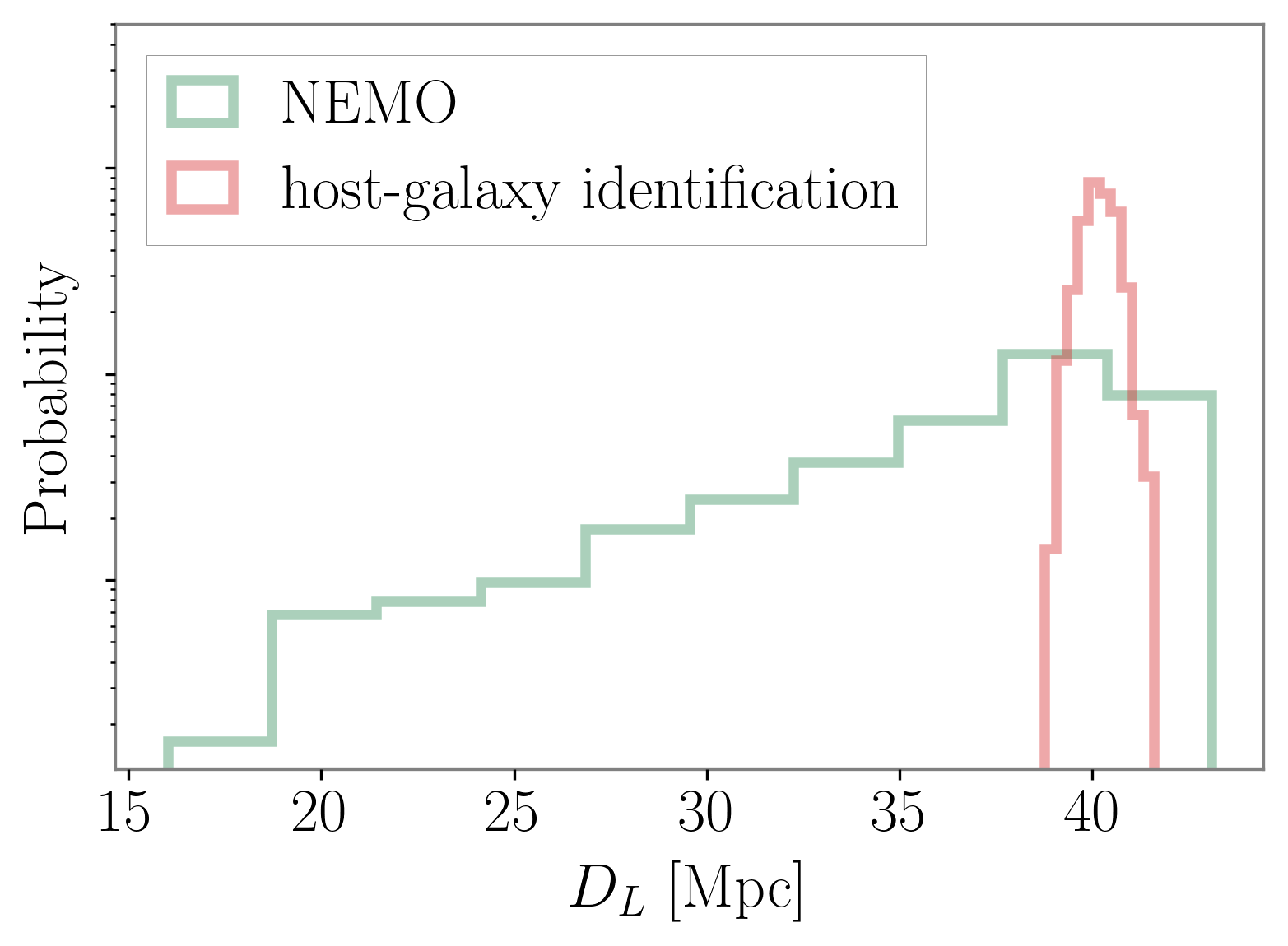}
    \caption{Posterior distribution for the luminosity distance $D_L$ for GW301116 from NEMO (green) and a host galaxy localisation (red) with a redshift uncertainty of $10^{-4}$.}
    \label{fig:dl} 
\end{figure}

We simulate an event coined GW301116, which is a face-on (inclination angle $\iota=0$) binary neutron star merger at luminosity distance $D_L=\unit[40]{Mpc}$ with component masses $m_1=\unit[1.5]{M_{\odot}}$ and $m_2=\unit[1.4]{M_{\odot}}$, dimensionless tidal deformabilities $\Lambda_1=400$ and $\Lambda_2=450$, dimensionless in-plane spins $\chi_{1,2}=0.02$, at ${\rm RA}=1.375$ and ${\rm Dec}=0.55$. We inject this signal into a single NEMO detector using the \verb!IMRPhenomPv2_NRTidal!~\citep{imrphenompv2_nrtidal} waveform, and recover the system parameters using the same waveform. We utilise \verb!Parallel Bilby!~\citep{bilby, bilby2, pbilby} with the \verb!dynesty! nested sampler~\citep{dynesty}. The sky map showing the $50\%$ and $90\%$ posterior credible intervals is shown in Fig.~\ref{fig:sky}, and the one-dimensional marginalised posteriors for the time of coalescence and luminosity distance are shown in green in Figs.~\ref{fig:tc} and~\ref{fig:dl}, respectively.

The event GW301116 being on axis and at $\unit[40]{Mpc}$ implies the prompt gamma-ray signal would be observed by any telescope not occulted by the Earth~\citeg{howell19}. We consider capabilities of future gamma-ray telescopes SVOM~\citep{svom}, and Theseus~\citep{Theseus}, which are all capable of localising the gamma-ray burst to $\lesssim 15$ arcmin precision~\citep{Theseus}. We thus model the sky-localisation posterior as a two-dimensional distribution mimicking current gamma-ray detectors such as \textit{Swift} using the \verb!GWCelery! package~\citep{GWCelery}. This sky map is shown as the inset in Fig.~\ref{fig:sky}.

A prompt gamma-ray detection also provides a strong constraint on the merger time, even accounting for the uncertainty in the delay between the merger and prompt emission signal. By the late 2020s/early 2030s, we estimate that the uncertainty on the merger time from the identification of a prompt gamma-ray signal will be $\mathcal{O}(\unit[2]{s})$. We therefore draw samples for the time of coalescence from a uniform distribution spanning \unit[2]{s} to derive a posterior distribution for $t_c$, which is shown as the purple curve in Fig.~\ref{fig:tc}.

An on-axis binary neutron star merger at $\unit[40]{Mpc}$ will also be detectable as a kilonova with observatories such as Vera Rubin~\citep{vera_rubin} and the Zwicky Transient Facility~\citep{ztf_paper}, and as an afterglow with various broadband electromagnetic telescopes. These observations will enable the identification of the host galaxy and redshift, as well as an indirect measurement of the merger time. The precision for sky localisation of such an event will therefore be sub-arcsecond~\citep{svom, vera_rubin, Theseus}. This is such a small localisation area by comparison to gravitational waves that it is not visible in Fig.~\ref{fig:sky}. We estimate that by the late 2020s/early 2030s, the error on the redshift will be $\Delta z\sim\mathcal{O}({10^{-4}})$. We note that the error on the redshift for the host galaxy of GW170817, NGC4993, is $\mathcal{O}({10^{-3}})$~\citep{Hjorth2017,abbott17_gw170817_Hubble}, and improvements in electromagnetic surveys and better understanding of systematic uncertainties like peculiar velocities will improve this limit~\citep{howlett20}. We further note that for on-axis events it may be difficult to disentangle the optical afterglow from the kilonova but we ignore this complication.
We assume a Gaussian posterior of width $\sigma_z=10^{-4}$ for the redshift, which we convert to luminosity distance assuming Planck 2018 cosmology~\citep{planck18}. The luminosity distance posterior is shown in Fig.~\ref{fig:dl} as the red curve. 

Using these various posterior distributions, we calculate the odds using Eqs. \ref{eq:odds} and \ref{eq:overlap} as derived in App.~\ref{sec:method}. For a jackpot event like GW301116, the Bayes factor is $4.3\times10^{11}$. 
To estimate the prior odds, we consider how many events with a short gamma-ray burst, afterglow, and kilonova we expect in the four-dimensional combined uncertainty region (i.e., $t_c$, $D_L$, RA, and Dec.) given the binary neutron star merger rate. For a merger rate of $\unit[320^{+490}_{-240}]{Gpc^{-3} yr^{-1}}$, and uncertainties on the coalescence time, luminosity distance, and sky position, we expect one merger every $\approx 12$ years in the four-dimensional uncertainty region, implying a prior odds of $9.8\times 10^{11}$. This implicitly assumes that every binary neutron star merger at $D_L = \unit[40]{Mpc}$ will produce a detectable prompt gamma-ray emission, afterglow or kilonova, which is true for an on-axis merger such as GW301116, but may not be for other mergers. We explore the effect of the inclination in Sec.~\ref{sec:inclination}. For prior odds of $9.8\times 10^{11}$, the odds is therefore $4.2\times 10^{23}$, indicating it is $4.2\times 10^{23}$ times more likely that the observations come from the same astrophysical event than they do not. We note that in more realistic scenarios, the prior odds will be dictated by the rate of false positives for a given observatory, the true astrophysical rate, and the potential for event misidentification. We elaborate on these more realistic prior odds calculations in the next section. We note that the above scenario of observing an on-axis event at $D_L = \unit[40]{Mpc}$ is going to be exceedingly rare. In the next sections we consider more common types of events.  

\subsection{Coincident detections as a function of distance}\label{sec:distance}
The expected detection rate of GW301116-like events that include coincident gravitational-wave, prompt emission, afterglow, and kilonovae observations as well as host-galaxy identification is far less than one per year. More realistically, we expect most multimessenger gravitational-wave events to be observed only with one of the aforementioned emissions and at varying distances. To wit, we repeat the calculation of Sec.~\ref{sec:170817} but with a number of events at different distances from $D_L=\unit[40]{Mpc}$ to $\unit[400]{Mpc}$. All other parameters and the calculation remain the same as detailed in the previous section. 

The afterglow and kilonova provide only an indirect measure on the time of coalescence $t_{c}$. The peak timescales for an on-axis merger are dictated by the deceleration of the relativistic jet for the afterglow, and the timescale for the expansion of the ejecta to balance the diffusion timescale for the kilonova. 
These peak timescales are $\mathcal{O}(\unit[\mathrm{1-2}]{s})$ and $\mathcal{O}(\unit[1]{d})$, respectively. By the late 2020s/early 2030s, we predict that advancements in cadence of optical telescopes and theoretical modelling of these phenomena will enable identification of afterglows and kilonovae within approximately $\unit[3]{hr}$ and $\unit[6]{hr}$ from the merger time, respectively. We emphasize that this is speculation, and realistic values will depend on detector design and survey choices too difficult to predict. Depending on the scenario used, we take this time uncertainty as the width of the posterior on the time of coalescence $t_{c}$.

The prior odds depend on the scenario being considered. For a jackpot event like GW170817, i.e., an event with an observation of a short gamma-ray burst, multi-wavelength afterglow and kilonova, the prior odds is the inverse of the number of events expected in the given four-dimensional combined uncertainty region given the binary neutron star merger rate itself. Given the low merger rate, this implies \textit{a priori} that if we see a confident short gamma-ray burst, kilonova, gravitational-wave signal, and afterglow in the same four-dimensional combined uncertainty region, it is overwhelmingly likely they are coincident (see previous section for the calculation). The same is not true for other scenarios. For example, the prior odds on a gamma-ray burst in coincidence with a gravitational-wave signal is dictated by the rate of triggers (astrophysical or otherwise) from gamma-ray telescopes. For Theseus, this is predicted to be $\unit[1-8]{week^{-1}}$~\citep{Theseus}, which dwarfs the astrophysical rate of detectable short gamma-ray bursts themselves. 

Optical telescopes will find many potential candidates for both kilonovae and afterglows. Even if these candidates are astrophysical, they could easily be mistaken as supernovae or afterglows of long gamma-ray bursts for example. A realistic rate of triggers in an observatory such as Vera Rubin is not known and will depend on the survey strategy~\citep{Setzer2019, Cowperthwaite2019}. We therefore take estimates from current wide-field optical telescopes, using particularly the Gravitational-Wave Optical Transient Observer~\citep[GOTO; ][]{goto}, which has recently announced results from a survey that searched for the optical afterglows of gamma-ray bursts, and had a trigger rate of $\approx \unit[1-20]{week^{-1}}$~\citep{Mong2021}. This is significantly larger than the astrophysical rate of kilonova-like or afterglow-like optical transients, and can be used to determine a conservative estimate of the prior odds for scenarios involving kilonovae and afterglows. We use these trigger estimates, alongside the rate of binary neutron star mergers and potential contaminants such as long gamma-ray bursts and their afterglows~\citep{Ghirlanda2021} and supernovae and fast-blue optical transients~\citep{Ho2020_koala}, to calculate the prior odds. 

We note that alongside binary neutron star mergers, some fraction of neutron star-black hole mergers are expected to produce a kilonova, short gamma-ray burst and afterglow. In fact, there may already be evidence of this contamination in the observed sample of short gamma-ray bursts~\citep{Siellez2016, Gompertz2020, Hamburg2020}. However, given the astrophysical rate of neutron star-black hole mergers is small~\citep{LIGOScientific:2021qlt}, ignoring contamination from neutron star-black hole mergers does not affect our result.

\begin{figure}
    \centering
    \includegraphics[width=0.5\textwidth]{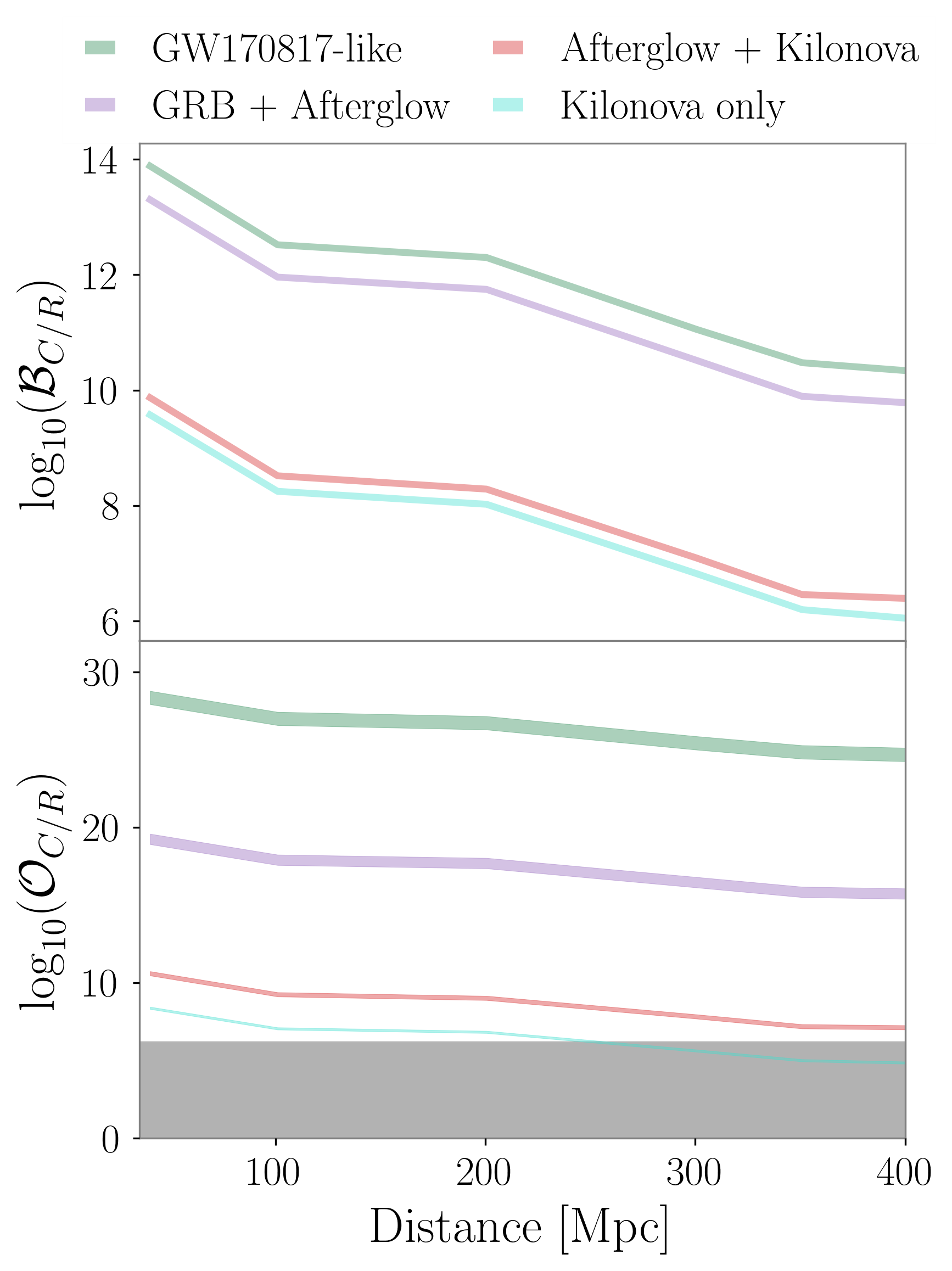}
    \caption{Bayes factor (top panel) and odds (bottom panel) as a function of distance for coincident detections of gravitational waves and various electromagnetic messengers. All events are on axis, with $\iota=0$.  For example, the turquoise curves show the Bayes factor and odds for coincident detections of gravitational waves with NEMO and a kilonova signal. The blue curve shows the coincident detection for a GW170817-like event; that is, the coincident detection of the prompt emission, broadband afterglow, and kilonova signal, as well as the gravitational-wave signal with NEMO. The width of the band represents the $90\%$ credible interval from marginalising over the prior odds.}
    \label{fig:BFOdds} 
\end{figure}

The green ``GW170817-like'' curve in Fig.~\ref{fig:BFOdds} shows the Bayes factor (top panel) and odds (bottom panel) as a function of distance for events where we detect the gravitational-wave signal as well as the prompt emission, broadband afterglow and kilonovae, as well as identify the host-galaxy. At all distances, the odds is overwhelmingly in favour of coincidence, going from $8.3\times10^{27}$ at $D_L=\unit[40]{Mpc}$, to $1.7\times10^{24}$ at $D_L=\unit[400]{Mpc}$, largely due to the strong a priori odds for a coincident event in that scenario.  The purple ``GRB + Afterglow'' curve shows the same, however assuming that the kilonova was not observed. The red ``Afterglow + kilonova'' curve assumes only the prompt emission is missed, while the turquoise ``Kilonova only'' curve assumes only the gravitational-wave and kilonova signal are detected (i.e., the prompt and afterglow are missed). 

The horizontal grey band in the bottom panel shows an odds of $\log_{10}\mathcal{O}_{C/R}=6$, corresponding approximately to a five-sigma threshold for the association of the gravitational-wave signal with the electromagnetic counterpart. Using this relatively arbitrary threshold, we see that all face on binary neutron star mergers will have confident coincidences between the gravitational-wave and electromagnetic counterpart, with the exception of signals that are only detected through their kilonova at distance $D_L\gtrsim\unit[250]{Mpc}$. We note that this threshold is somewhat arbitrary, and deriving a realistic threshold requires understanding the distribution of $\log_{10}\mathcal{O}_{C/R}$.

\subsection{The effect of inclination angle}\label{sec:inclination}
The previous two sections assume the gravitational-wave and electromagnetic signal(s) are observed, and quantifies the confidence we could get that those signals originate from the same source. Not all electromagnetic signals will be detectable, in part due to the inclination angle of the source implying, for example, the gamma-ray burst jet is pointed away from Earth. To understand this, we look at the number of detectable gamma-ray burst, afterglow, kilonovae, and gravitational-wave sources as a function of both distance and inclination angle.

At fixed values of distance, we create a population of neutron star mergers that are isotropically distributed in inclination angle (i.e., uniform in $\cos\iota$) with the same intrinsic parameters as described in Sec.~\ref{sec:2030s}.
We assume the same jet structure and kilonova models as described in the previous sections. 
For each event, we calculate whether the signal is detectable with each messenger and, if detectable, whether any two or more messengers can be confidently associated. We marginalise over systematic uncertainties due to the unknown equation of state, jet structure and prior odds discussed in Sec.\ref{sec:EM} and \ref{sec:assoc}, which directly affect the expected kilonova, gamma-ray burst signature and odds calculation, respectively.

\begin{figure*}
    \centering
    \includegraphics[width=0.7\textwidth]{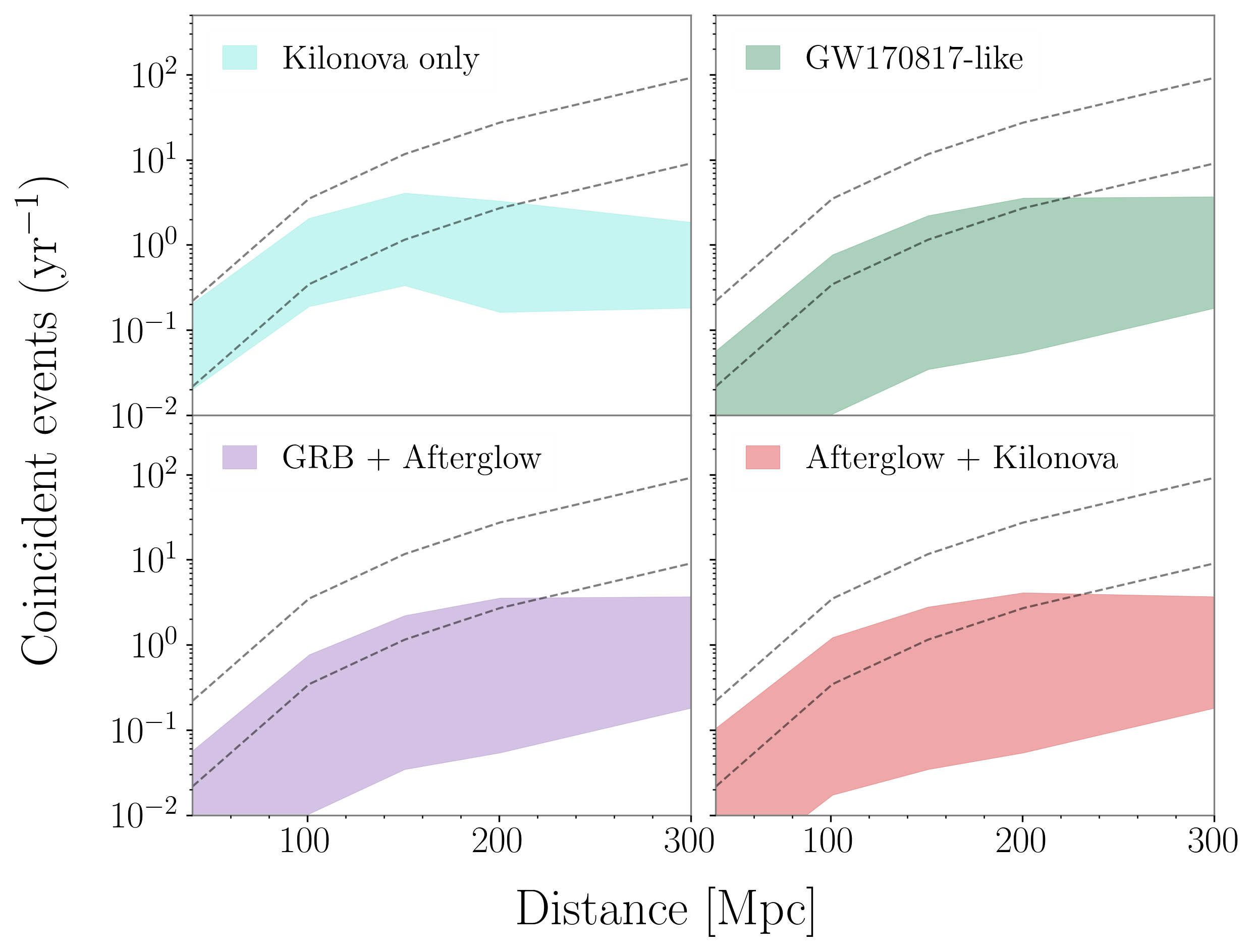}
    \caption{The number of coincident events as a function of distance with $\log_{10}\mathcal{O_{C/R}} \gtrsim 6$ corresponding to five sigma confidence. For example, the top left panel shows the total rate of confident coincident detections that can be made between gravitational-wave and kilonova signals. The dashed curves indicate the $90\%$ credible interval of the binary neutron star merger rate, and the coloured band represents the $90\%$ credible interval from marginalising over the model uncertainty and prior odds.}
    \label{fig:coincidentevents} 
\end{figure*}

In Fig.~\ref{fig:coincidentevents}, we show the total coincident detection rate per year as a function of luminosity distance. As an example, the top left panel shows in the turquoise-coloured band the total detection rate for which the gravitational-wave signal and kilonova can be confidently identified (above a $5-\sigma$ threshold) as coming from the same source, marginalising over all previously-mentioned uncertainties, as well as the uncertainty in the overall merger rate. This coloured band can be contrasted with the total binary neutron star merger rate, where the 90\% confidence intervals are shown as the dashed curves in each panel. 

One can inspect the curves in Fig.~\ref{fig:coincidentevents} to derive an overall merger fraction $\mathcal{F}$ of mergers that will be detected with gravitational waves and various electromagnetic messengers, as well as the overall rate $\mathcal{R}$ of detections. These numbers are provided in Tab.~\ref{tab:rates}. These numbers show $\coincident{}$ (or $\approx 40\%$) of binary neutron star mergers per year will have confident associations between their independently-identified electromagnetic signatures and gravitational-wave signal with only a NEMO detector. 

We note that the above numbers, in particular the fraction of binary neutron star mergers with any confident association, should be taken with some caution. These numbers are sensitive to three assumptions: 1) The true distribution of $\log_{10}\mathcal{O}_{C/R}$ is distributed in a way such that our threshold for a confident association is too weak. 2) The modelling assumptions and priors for the electromagnetic models are too optimistic. 3) The fraction $\mathcal{F}$ represents the fraction of mergers detectable with both an electromagnetic telescope and with NEMO that can also be confidently associated with one another. However, whether a given electromagnetic telescope can actually detect a counterpart (even if it is above the threshold for detection) will depend on the survey strategy and specifics of the detector. These strategies and detector specifics are too difficult to predict for observatories that do not yet exist, and a calculation of the true distribution of $\log_{10}\mathcal{O}_{C/R}$ is not computationally feasible as it requires simulating the complete population multiple times to create a distribution of $\log_{10}\mathcal{O}_{C/R}$ for each binary neutron star merger. However, both of these factors will likely be better understood in the near future, accompanied by a better understanding of electromagnetic models.

\begin{table*}[h!]
    \centering
    \begin{tabular}{|c| c c c c |}
        \hline
         &      prompt      & prompt    &           &  \\
         &      afterglow   & afterglow & afterglow & \\
         &      kilonova    &           & kilonova  & kilonova \\
        \hline
        $\mathcal{F}$ ($<\unit[300]{Mpc}$) & $0.096^{+0.072}_{-0.068}$ & $0.096^{+0.072}_{-0.068}$ & $0.13^{+0.13}_{-0.09}$ & $0.38^{+0.3}_{-0.01}$ \\
        $\mathcal{R}$ (${\rm yr}^{-1}$) & $\jackpot{}$ & $\grbag{}$ & $\knag{}$ & $\knonly{}$ \\
        \hline
    \end{tabular}
    \vspace{0.5em}
    \caption{Expected fraction $\mathcal{F}$ and rate $\mathcal{R}$ of coincident gravitational-wave and electromagnetic observations using a single NEMO-like detector. The fraction measures the fraction of binary neutron star mergers occurring at a luminosity distance of less than $\unit[300]{Mpc}$ that have an identified electromagnetic counterpart.  For example, the first column shows the fraction (first row) and number (second row) of expected mergers for which the gravitational-wave detection can be confidently associated with observations of both prompt emission, afterglow, and kilonova. We note that all systems detected through their prompt emission will also be detected as afterglows. For comparison, a single NEMO-like detector will detect the gravitational-wave signal from $\mathcal{F}=0.45$ mergers out to $\unit[300]{Mpc}$, corresponding to an event rate of $\unit[16^{+26}_{-12}]{yr^{-1}}$.}
    \label{tab:rates}
\end{table*}

\section{Conclusions}\label{sec:conclusion}
The future of gravitational-wave astronomy relies on sequential technological and infrastructure upgrades, as well as the development of objectively expensive new facilities (albeit far less than some space-based infrared telescopes).
One plausible scenario sees a period of time for which a designated kilohertz gravitational-wave detector like the proposed NEMO instrument~\citep{NEMO} is the only operational ground-based observatory. In this work, we show that such an instrument has significant scientific utility on its own due to the multimessenger capabilities expected in the same era. For example, in operation with the Vera Rubin observatory~\citep{vera_rubin} and Theseus~\citep{Theseus}, we show that approximately $40\%$ of binary neutron star mergers out to $\unit[300]{Mpc}$ can be confidently identified as multimessenger coincident observations. Given current estimates for the local merger rate of neutron stars~\citep{Abbott2021_03acatalog}, this implies a total number of $\coincident{}$ multimessenger detections per year. 

Coincident multimessenger observations of binary neutron star mergers will enable several precise tests into the nature of nuclear matter. The $\gtrsim\unit[500]{Hz}$ gravitational-wave sensitivity enables simultaneous measurements of the neutron star's masses and tidal deformability, which can be further constrained when coupled with kilonovae observations that inform the ejecta mass, and hence the progenitor's mass ratio and complementary information about the equation of state~\citeg{metzger17}. For nearby mergers~\citep[see][]{NEMO} this will be combined with measurements of gravitational waves from the hot post-merger remnant, which can further inform the equation of state~\citeg{bauswein19}, including the potential to discern temperature-dependent phase transitions~\citep{bauswein16}. Combining gravitational-wave information from the inspiral and post-merger remnant with constraints on the jet-launching timescale from timing of the prompt emission~\citep{ren20_delay, beniamini20_delay} can constrain the maximum mass of hot and cold neutron stars~\citeg{Chatziioannou2020, sarin21_review}; also a key indicator of the nuclear equation of state. This information can also be determined from complementary observations of, for example, the colour and energetics of the kilonova~\citep{margalit19} and the duration of x-ray plateaus from surviving supramassive neutron stars~\citep{rowlinson13, ravi14, sarin20}.

Associating gamma-ray bursts and kilonovae with gravitational-wave signals will also enrich our understanding of the physics driving these enigmatic transients. In particular, constraints on the delay time between merger and prompt emission will deepen our understanding of ultra-relativistic jet launching and propagation~\citep{ren20_delay}. This will also shed insight into the prompt emission mechanism. Detection of post-merger gravitational waves combined with prompt emission observations will provide a smoking-gun observation into the contentious issue of whether a black-hole central engine is required to launch an ultra-relativistic jet capable of generating short gamma-ray burst emission. This will also provide an opportunity to understand the impact of the remnant on r-process enrichment~\citeg{perego14, metzger17, bernuzzi20}. Coincident multimessenger observations will also improve our understanding of the jet structure and beaming of gamma-ray burst jets~\citeg{beniamini20a, Biscoveanu2020}.  

The capabilities of survey telescopes in the 2030s will enable host-galaxy identification of neutron star mergers, and hence precise redshift measurements~\citeg{Howlett2017}. Required peculiar-velocity measurements will be enabled by next-generation spectroscopic surveys~\citeg{palmese19,howlett20}, and possibly percent-level measurements of the Hubble constant with single events~\citep{coughlin20,calderonbustillo21}.

Dedicated kilohertz gravitational-wave observatories have the power to produce valuable science \textit{without} the requirement of living in a global, heterogeneous network of other gravitational-wave observatories, provided the capabilities of electromagnetic telescopes do not get worse. In this paper, we show this is true only considering the guaranteed science case of binary neutron star mergers. Neutron star-black hole binaries, millisecond pulsars, and supernovae are all targets for NEMO-like observatories that would provide rich multimessenger information about the physics and astrophysics of these extreme objects. However, the unknown neutron star-black hole tidal-disruption rate, the unknown ellipticity of rapidly-rotating isolated neutron stars, and the unknown amplitude of gravitational waves from supernovae prohibits us from making concrete statements on the likelihood of coincident detections. Regardless, we know that the science outcomes from such events would likely be bountiful.

\begin{acknowledgements}
We are grateful to David Ottaway and Eric Howell for their thoughtful comments on the manuscript. This work was supported through Australian Research Council (ARC) Centre of Excellence CE170100004, ARC Future Fellowship FT160100112, and ARC Discovery Project DP180103155. This work was performed on the OzSTAR national facility at Swinburne University of Technology. The OzSTAR program receives funding in part from the Astronomy National Collaborative Research Infrastructure Strategy (NCRIS) allocation provided by the Australian Government.
\end{acknowledgements}

\appendix

\section{Nitty-gritty formalism}\label{sec:method}
We wish to determine the probability that observed electromagnetic and gravitational-wave signals are coming from the same source. This involves two questions: 1) are the two events coincident in space and time, and 2) is it physically possible/probable that the signals were generated in the same event. In a Bayesian sense, the former question is quantified by the Bayes factor, while the latter is quantified by the prior odds. Together, they enable us to calculate the odds that any two or more pieces of data were generated by the same source.


To determine whether two or more independent observations are coincident (i.e., come from the same physical source), we generalise the Bayesian formalism of~\citet{Ashton2018} from two to $N$ coincident data sets. We calculate the \textit{odds}~$\mathcal{O}_{C/R}$~\citep{Ashton2018, Ashton2020_gw190521}, where $C$ refers to the hypothesis that the two independent detections share a common origin and $R$ refers to the hypothesis that they are random. This is given by
\begin{equation}
    \mathcal{O}_{C / R} \equiv \mathcal{B}_{C/R}\pi_{C / R},\label{eq:odds}
\end{equation}
where 
\begin{equation}
    \mathcal{B}_{C/R}\equiv \frac{P\left(C \mid d_{1}, d_{2}, \ldots, d_{N}, I\right)}{P\left(R \mid d_{1}, d_{2}, \ldots, d_{N}, I\right)}, 
\end{equation}
is the Bayes factor, ${P\left(C \mid d_{1}, d_{2}, \ldots,d_{N}, I\right)}$ is the probability of a common hypothesis conditional on data sets $d_{1}, d_{2}, \ldots,d_{N}$, while the denominator refers to the random hypothesis conditioned on the same data. The second term in Eq.~\ref{eq:odds}, $\pi_{C / R}$ is the \textit{a priori} probability of a coincident event, which conservatively can be estimated as one over the number of expected events in the given overlapping region. 

For a given gravitational-wave and electromagnetic transient to share a common origin, they must have occurred at the same location in the sky, at the same distance, and within some coincident time-span as specified by theoretical models. 
This implies they must have the same sky-location parameters $\Omega$, same luminosity distance $D_L$, and same inferred time of coalescence $t_{c}$. 
In the case of observations with a common source, some other parameters will also be the same (e.g., source inclination). However, given theoretical limitations in current electromagnetic models, particularly in connecting to the progenitor parameters measured by the gravitational-wave signal, we ignore these common parameters. The confidence gained by including these additional parameters is proportional to the information gained on these parameters relative to the prior. 

Following the framework from~\citet{Ashton2018} the numerator of the Bayes factor is, 
\begin{equation}
\begin{split}
    P\Big(C \mid &d_{1}, d_{2}, \ldots,d_{N}, I\Big) \\ 
    &=\int P\left(C, \theta \mid d_{1}, d_{2}, \ldots ,d_{N}, I\right)\pi(\theta \mid C)~d \theta.
\end{split}
\label{eq:bayesnumerator}
\end{equation}
Here, $\theta$ is the set of parameters common to all data sets, $\pi(\theta \mid C)$ is the common-hypothesis prior on $\theta$, and
\begin{equation}
P\left(C, \theta \mid d_{1}, d_{2}, \ldots,d_{N}, I\right) = \prod\limits_{i=1}^{N} \frac{P\left(d_{i} \mid C\right) P\left(\theta \mid d_{i}, C\right)}{P\left(\theta \mid C\right)}.
\label{eq:genbayes}
\end{equation}
Substituting Eq.~\ref{eq:genbayes} into Eq.~\ref{eq:bayesnumerator} gives, 
\begin{equation}
\mathcal{O}_{C/R} = \pi_{C/R}\mathcal{I}_{\theta},
\label{eq:genoverlap}
\end{equation}
where $\mathcal{I}_\theta$ is the posterior overlap integral for the parameter $\theta$ generalised for $N$ different data sets 
\begin{equation}
    \mathcal{I}_{\theta}=\int \prod_{i=1}^{N} p\left(\theta \mid d_{i}, C\right)\pi(\theta \mid C)^{1-N} d \theta. \label{eq:overlap}
\end{equation}

We now move from the generalised abstract formalism described above to our specific study. For example, we consider the scenario of three different data sets: a gravitational-wave signal, kilonova, and afterglow. The overlap integral can be written as 
\begin{equation}
\mathcal{I}_{\theta}=\int \frac{p\left(\theta \mid d_{\mathrm{gw}}, C\right) p\left(\theta \mid d_{\mathrm{kn}}, C\right) p\left(\theta \mid d_{\mathrm{ag}}, C\right)}{\pi(\theta \mid C)^{2}} d \theta,
\end{equation}
where $d_{\mathrm{gw}}$, $d_{\mathrm{kn}}$, and $d_{\mathrm{ag}}$ are the gravitational-wave, kilonova and afterglow data, respectively. 
The odds given our common parameters $\theta=\{D_{L},\,\Omega,\,t_{c}\}$ is
\begin{equation}
\mathcal{O}_{C/R} \approx \pi_{C/R}\mathcal{I}_{D_L}\mathcal{I}_{\Omega}\mathcal{I}_{t_{c}},\label{eq:partialodds}
\end{equation}

In main body of the Paper we discuss what constraints a detection of kilonovae, prompt gamma-ray and afterglow emission and a gravitational-wave signal detected by a kHz-band gravitational-wave observatory can provide on these joint parameters. We use the framework described above to determine how confidently we can associate a given electromagnetic counterpart to the gravitational-wave signal. We note that since these transients are independently identified, we do not rely on low-latency estimates to calculate the odds and associate events. 

\section{Priors}\label{appendix:priors}
Here we list the priors used to model the prompt emission, afterglow, and kilonova signal from a binary neutron star merger. These priors are motivated by analysis of the afterglow of GRB170817A~\citep{afterglowpy} and the kilonova AT2017gfo~\citep{Coughlin2017}. The priors are displayed in Tab.~\ref{tab:priors}.
\begin{table*}
\centering
 \caption{Parameters used in our modelling of the electromagnetic counterparts to binary neutron star mergers along with a brief description and the prior.}
 \label{tab:priors}
 \begin{tabular}{lcc}
  \hline
  Parameter [unit] & Description & Prior \\
  \hline
$\Gamma_0$ & initial Lorentz factor & $\textrm{Uniform}[100,1000]$ \\
$\log_{10} (E_{\textrm{iso}}/\rm{erg}) $ & isotropic-equivalent energy & $\textrm{Uniform}[52,54]$\\
$\theta_{\textrm{core}}$ [rad] & half-width of jet core& $\textrm{Uniform}[0.1,0.2]$ \\
$\log_{10} (n_{\textrm{ism}}/\unit{cm^{-3}})$ & number density of ISM & $\textrm{Uniform}[-4,1]$\\
$p$ & electron distribution power-law index & $\textrm{Uniform}[2,3]$\\
$\log_{10}\epsilon_{e}$ & thermal energy fraction in electrons& $\textrm{Uniform}[-3,0]$\\
$\log_{10}\epsilon_{b}$ & thermal energy fraction in magnetic field & $\textrm{Uniform}[-4,0]$\\
$\xi_{N}$& fraction of accelerated electrons & $\textrm{Uniform}[0,1]$\\
$\log_{10} p_c $& neutron star central pressure & $\textrm{Uniform}[32.6,33.6]$\\
$\Gamma_{1, 2, 3}$ & polytropic indices & $\textrm{Uniform}[1.1,4.5]$\\
$\beta$ & ejecta velocity distribution power-law slope & $\textrm{Uniform}[3,5]$\\
$\kappa$ & kilonova ejecta opacity & $\textrm{Uniform}[1,10]$\\
$M_{\rm{TOV}}$ [$M_{\odot}$] & Maximum non-rotating neutron star mass & $\textrm{Uniform}[2.2,2.4]$\\
  \hline
 \end{tabular}
\end{table*}

\bibliographystyle{pasa-mnras}
\bibliography{Bib}

\end{document}